# Quantifying the Rate Performance of Potassium-Ion Batteries

Caolin Ua Tuiscint and Jonathan N Coleman

*School of Physics, CRANN & AMBER Research Centres, Trinity College Dublin, Dublin 2, Ireland*

*colemaj@tcd.ie (Jonathan N. Coleman); Tel: +353 (0) 1 8963859.

**Abstract**

In battery research and industry, lithium-ion batteries are dominant due to their legacy of research and high energy density. However, due to the high demand for batteries and limited resources of lithium, lithium prices are high and increase the cost of batteries. As a result, potassium-ion batteries have in recent decades begun being researched, due to potassium's greater abundance and potential affordability. Although, due to their infancy they require key insights for greater development. Recently, a set of methods was established which enable rigorous analysis of electrode rate performance. In this work they are applied, published rate-performance data for a wide range of potassium ion batteries has been analysed to determine the success of the key model of these methods, the opportunities within potassium-ion batteries, and how these batteries compare to previous batteries is assessed using these methods. Using common specific capacity(mAh/g) versus charge/discharge rate curves, parameters which quantify performance were extracted, the success of the model was determined by the effectiveness of these parameters to be extracted, and equations which relate these performance parameters to electrode properties were used to analyse potassium-ion batteries. Analysis of the key model found it to effectively fit the potassium-ion battery rate-performance data. Additionally, comparison of potassium ion batteries found that while specific capacity lagged that of lithium-ion batteries and sodium-ion batteries, despite the large atomic size and mass of potassium, the upper limit of rate performance for potassium-ion batteries was found to be superior to that of LIB, SIB, or their 2D material electrodes.

**Introduction**

Batteries are a unique solution that can enable the decarbonisation of the majority of global emissions, allowing industrial energy use, residential energy use, and transport to transition to sustainable sources of energy. [1-3] The focus of most battery research and production goes into lithium-ion batteries, due to its promise in terms of energy density as the lightest alkali metal. [4-6] However, due to cost and limited reserves, research in alternative batteries has become more prominent. [7-9] Due to the demand for batteries and the limited resources of lithium the price of LIBs is high. [8] Additionally, the reserves of lithium do not meet expected future demand. [8-11]

So, due to its abundance and its place after lithium in the alkali metals group, sodium was later developed as a battery material. [12] While now potassium is being developed, as it also follows both lithium and sodium in the alkali metals group. [13] Potassium is more abundant than lithium, and additionally it has a greater electrochemical potential than that of sodium, making its batteries too potentially more affordable. [13]

However, LIBs today greatly benefit from their legacy of development, while sodium-ion batteries(SIBs) and potassium-ion batteries(PIBs) have only just begun to be developed in the last couple of decades. [12-14] So, much of the initial data which provided researchers the insight on promising directions of research has not yet been developed for these new batteries. While, issues that had been resolved through research for LIBs will now once more have to be researched for SIBs and PIBs. [11-16]

This work hopes that by reviewing previous experimental research and conducting a meta-analysis, using some recently established tools, insights can be provided that can bridge this gap. Recently, our group established a method of quantifying the rate performance of batteries. [17, 18] These methods could provide these critical early insights into the performance of these new batteries. Our methods have already proved useful in the analysis of 2D material electrodes, diagnosing issues relating to their rate performance. [18] They also proved useful in the analysis of lithium-ion and sodium-ion batteries, which is why in terms of alternative batteries this work focuses on potassium-ion batteries.

Our method accurately models the relationship between specific capacity (mAh/g), and charge/discharge rate($h^{-1}$). When this model is used to fit experimental data of specific capacity versus rate, it produces a set of characteristic parameters which denote an electrode's specific capacity and charge/discharge rate performance. And, by quantifying these performance characteristics for a cohort of these new batteries, these cohorts can be assessed and compared directly to one another. In fact, these parameters were also established as being directly related to the performance of different processes and the physical properties of the battery. This allows analysis to determine the source of issues and methods of improving these characteristic performance parameters. So by conducting this analysis these new batteries can be investigated to determine what is required to be improved and how they can be, while their performance can be compared to LIBs to view their drawbacks and opportunities.

**Results & Discussion**

**Result of Fitting Capacity versus Rate Data**

It is well-known that the capacity of battery electrodes decreases as their charge/discharge rate increases (see Figure 1). Using this data to quantify rate performance has, in the past, not been simple. Recently, [17] we proposed a semi-empirical equation, which fits this capacity-rate data producing three characteristic fitted parameters:

$$\frac{Q}{M} = Q_M[1 - (R\tau)^n(1 - e^{-(R\tau)^{-n}})] \tag{1}$$

where Q/M is the measured specific capacity (mAh/g); R is rate defined through specific current (I/M), and the measured specific capacity at that given rate, as R = (I/M)/(Q/M). Making R representative of the actual charge/discharge time. The three characteristic fitted parameters are: $Q_m$, $\tau$, and $n$. $Q_M$ is the low-rate capacity, representing the greatest value specific capacity will reach. $\tau$ is the characteristic time constant, representing the inverse of the rate, 1/R, where capacity has fallen by 1/e. This means $\tau$ marks a reference point within each electrode's capacity decay, the point at which high rate decay has begun, which occurs at a unique rate for each electrode. This parameter is particularly important, with low values indicating good rate performance. While n is the exponent denoting the rate or slope of capacity decay at high rate. Low values of n indicate slower decay and good rate performance. Our previous work found that values of n about 1 were associated with electrodes with electrical rate limitations, while values of n about 0.5 were associated with electrodes with diffusion rate limitations.

The objective of this work is to assess the effectiveness of our methods on PIBs, in addition to assessing the rate performance of PIBs and comparing their performance to battery cohorts previously assessed through the same methods. Our literature search successfully identified 53 appropriate electrode datasets within 31 research papers presenting electrode data for PIBs [19-49]. The fitting of published capacity versus rate datasets was very successful; given sufficient low and high rate data all capacity versus rate datasets could be effectively fit to Equation 2. Figure 1 demonstrates some of these fitted datasets, plotting specific capacity (mAh/g) versus rate ($h^{-1}$), these fits show successful fitting at both low and high rate, effectively representing the data.

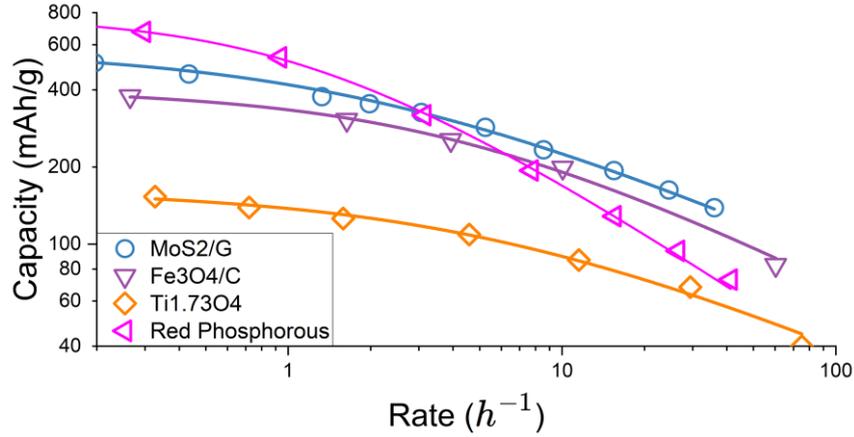

*Figure 1. Examples of fitting done on capacity versus rate data from PIB electrodes. [24, 25, 27, 48]*

The datasets collected stand as a good representative sample of PIB performance and this provides a great opportunity to understand the performance of PIBs. The availability of this data allows us to understand the range of specific capacity and rate performance, as well as the limitations of these batteries. To investigate the range of these fitted parameters, and their dependency to one another, Figure 2 shows plots of (A) n versus $Q_M$(mAh/g), (B) n versus τ(h), and (C) τ(h) versus $Q_M$ (mAh/g).

These plots include information on the material of the electrodes which are categorised into six groups: TMD, transition metal dichalcogenide materials including $WS_2$, $MoS_2$, $MoS_2$/Graphene, $CoSe_2$//g@NC, $CoSe_2 - FeSe_2$/g@NC, $FeSe_2$//g@NC, and $CoSSe - C$; Other Metal Chalcogenides materials including $Co_{1.67}Te_2$, $Sb_2Se_3$, and $Co_{0.85}Se$; Oxide and Hydroxide materials including $BiSbO_4$, $Ti_{1.73}O_4$, $Co_3O_4$/Mxene, $K_{0.4}Fe_{0.1}Mn_{0.8}Ti_{0.1}O_2$, PTCDA, and Magnetite; Carbon materials including Graphite, Graphene, and Soft Carbon; Phosphorous materials including Black and Red Phosphorous; As well as "Other" materials including Sb, SnSb, S, $Bi - AQ_{26}DS$, $CuP_2$, and $Fe_7(CN)_{18}$.

Additionally, within two research papers authors created multiples of the same electrode with varied thicknesses. In Figure 2 these datapoints are highlighted with hollow interiors. As the two papers used materials from two separate categories, it is easy to distinguish one from the other, one paper used an Other Metal Chalcogenides material and the other used an Oxide or Hydroxide material.

The distribution of values of $Q_M$ and $\tau$ are lognormal distributions, where the ranges for these parameters are 90 mAh/g < $Q_M$ < 1100 mAh/g and $1 \times 10^{-4}$ h < τ < 2 h, and the mean values of $\ln(Q_M)$ and $\ln(\tau)$ are 349.5 mAh/g and 0.0382 h respectively. The distribution of n values remain within a range of 0.25 < n < 1.5, where values are centred around two nodes, 0.45 and 0.87.

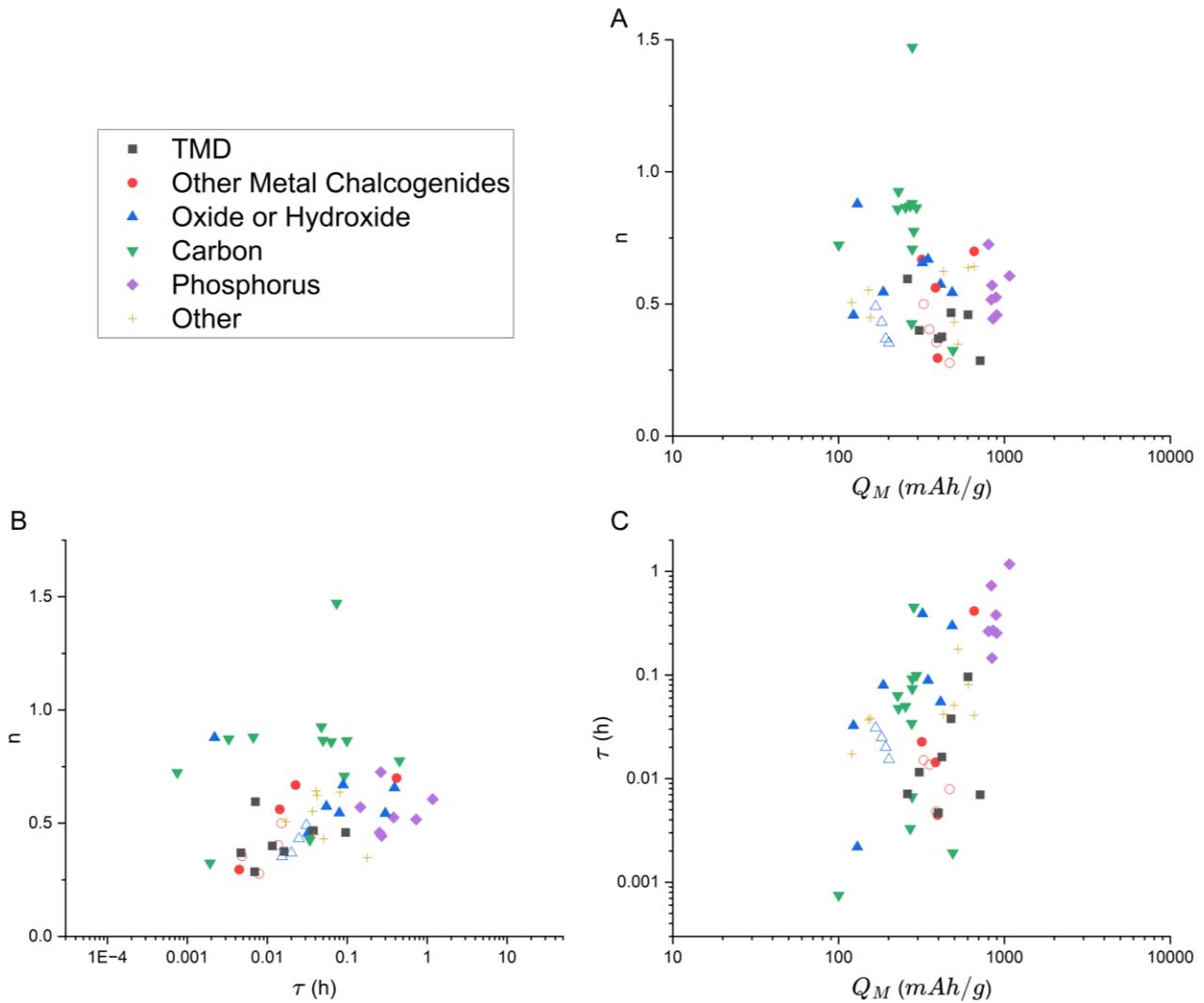

*Figure 2. Parameters obtained from fitting, produced by fitting 53 electrode specific capacity vs current/discharge rate data sets found within literature. This data is representative of 29 different active materials used for PIBs. Separated into the categories: Transition metal di-chalcogenides, other metal Chalcogenides, oxides or hydroxides, carbons, phosphorous and other miscellaneous materials. Each category is given its own symbol and colour as seen in the legend. In these plots (A-C) the fitted parameters, $Q_M$ (mAh/g), $\tau$ ($h^{-1}$), and n, are plotted against each other in three combinations. [19-49]*

For the distribution of $Q_M$, the representative range of maximum specific capacities reveals that PIB electrodes do demonstrate similar specific capacities to their LIB counterparts, although these competitors outmatch them in terms of greatest and average specific capacity.

The specific capacity range of LIB and SIB – 2D Materials is 200 mAh/g < $Q_M$ < 2000 mAh/g [17, 18]. The capacity of PIBs is likely impacted by the infancy of their research, as research into PIBs only began in the 2000s, while LIB research began in the 1960s [13, 14].

For the distribution of n, the reason it shows values centred around two points is likely representative of the factors that impact values of n, as mentioned previously. Values centred around 0.45 represent electrodes that have very little or no electrical limitation, while values centred around 0.87 represent electrodes heavily effected by electrical limitations. Although electrodes demonstrate electrical rate limitations, previous work has shown the role of conductive additive in eliminating these limitations[17], and improving all rate associated parameters. As most electrodes demonstrate values of n greater than 0.5, most electrodes would benefit from the addition of more conductive additive.

Additionally, many electrodes display values of n < 0.5. Given the known rate limiting characteristics at 0.5 and 1.0, there was reasonable concern that this may indicate an additional unknown limitation. Thus these low values were investigated, although it was found that these values were due to a lack of high rate data for these electrodes, which compromised the fitting of n. This analysis is shown in the discussion of Figure S1 within the supplementary information.

Assessing the relationships between the fitted parameters, Figure 2C shows a strong inversely proportional relationship between rate performance and capacity performance, $\tau$ and $Q_M$. Whereas, plots of (A) $n$ versus $Q_M$ and (B) n versus $\tau$ display no clear relationship. The relationship between n and $\tau$ may have some correlation but it is better described as scatter centred around a point. To compare material performance, it is evident phosphorus holds the greatest specific capacity performance while maintaining worse rate performance than all other materials, although its rate is better than the overall correlation between $Q_m$ and $\tau$. TMDs and Other Metal Chalcogenides also maintain a relatively high specific capacity compared to the overall correlation and to peers with similar values of $\tau$. Carbonous materials have mixed performance, some showing capacity well above theoretical limits with rate performance far better than peers at similar capacities. However the majority of carbonous materials show poorer rate performance compared to peers while having low to median specific capacities.

**Figure of Merit for Rate Performance**

In addition to producing a model for fitting capacity decay which provides parameters of electrode performance[17], we determined an equation linking $\tau$ to an electrode's physical properties, which allowed us to produce a figure of merit for rate performance.

By combining the mechanistic factors - the characteristic time associated with ion diffusion, the RC charging time of the electrode, and the time scale associated with the electrochemical reaction - we produced the equation:

$$\tau = L_E^2 \left[ \frac{C_{V,eff}}{2\sigma_E} + \frac{C_{V,eff}}{2\sigma_{P,E}} + \frac{1}{D_{P,E}} \right] + L_E \left[ \frac{L_S C_{V,eff}}{\sigma_{P,S}} \right] + \left[ \frac{L_S^2}{D_{P,S}} + \frac{L_{AM}^2}{D_{AM}} + t_C \right] \quad (2)$$

Where $C_{V,eff}$ is the electrode's effective volumetric capacitance, $\sigma_E$ is the out-of-plane electrical conductivity of the electrode material and the overall (anion and cation) conductivity within the bulk electrolyte, $\sigma_{P,E}$ and $\sigma_{P,S}$ are the ionic conductivity within the pores of the electrode and the separator, $D_{AM}$ is the solid-state ionic diffusion coefficient in the active material, $D_{P,E}$ and $D_{P,S}$ are the ionic diffusion coefficient in the pores of the electrode and separator, $L_{AM}$ is the length associated with solid-state diffusion within the active material particles, $L_E$ and $L_S$ are the thickness of the electrode and separator, and $t_c$ is the time scale associated with the electrochemical reaction once the electron and ion combine at the active particle. The formulation of this equation and explanation of its parameters have been described in greater detail previously. [17] The applicability of this equation has been proven, as it has been shown to accurately describe a wide range of experimental data.

Additionally, recently we proposed a more simplified and improved model. [18] We asserted that terms 1 and 7 in our equation are not important under all circumstances and can be neglected. We noted that our empirical observations show $C_{V,eff}$ is directly proportional to the volumetric capacity of an electrode, $Q_V$, where $C_{V,eff} = 28Q_V$. Additionally, we note that in porous systems diffusivity and ionic conductivity tend to be reduced by a factor $f$, a tortuosity factor, from the same measure in the bulk electrolyte. For example within the electrode $f = D_{P,E}/D_{BL} = \sigma_{P,E}/\sigma_{BL}$, where $D_{P,E}$ and $\sigma_{P,E}$ are the diffusion coefficient and conductivity of ions in the electrode's pores. Additionally we approximate, in the separator $f_s \approx P_s$, separator porosity. By combining these new additions we yield the equation:

$$\frac{\tau}{L_E^2} \approx \left[ \frac{14 Q_V}{\sigma_{BL} f} + \frac{1}{D_{BL} f} + \frac{28 Q_V L_S/L_E}{P_S \sigma_{BL}} + \frac{L_S^2/L_E^2}{P_S D_{BL}} + \frac{L_{AM}^2/L_E^2}{D_{AM}} \right] \quad (3)$$

From these equations, we can tell that $\tau$ is a quadratic function of $L_E$. However, upon assessing $\tau$ in LIBs and SIBs, in our previous works, it was evident that at significant thicknesses $\tau$ approximately scales with $L_E^2$, $\tau \propto L_E^2$, as the values of the other terms become insignificant in comparison. Due to this relationship with $L_E$, $\tau$ cannot be appropriately compared between different electrodes to determine which has superior rate performance, as the different electrodes will likely have significantly differing electrode thicknesses. However, having determined this relationship between $\tau$ and $L_E$, it allowed us to produce a figure of merit which can allow the comparison of rate performance of electrodes:

$$Figure\ of\ merit\ for\ rate\ performance = \Theta = \frac{\tau}{L_E^2}$$

This allows experimental researchers creating electrodes to calculate a metric for rate performance and determine how the performance of their electrode compares to others, determining value and possible outlying issues. Although, for our purposes this also allows us to compare cohorts of electrodes such as a representative sample of PIBs to that of LIBs and SIBs. Our previous work studied two cohorts one cohort was of LIB and SIB electrodes – all material types and the other was LIB and SIB electrodes – of a strictly 2D active material morphology, throughout this paper the results of these groups will be used to compare against PIBs. To compare the rate performance of these cohorts, Figure 3(A-C) shows the distributions for PIBs (A), LIB and SIB – All Materials (B), and LIB and SIB – 2D Materials (C), plotting the FoM as $Log(L_E^2/\tau)$ $[Log(m^2/s)]$ along the x axes and the amount of datasets, the count, along the y axes. All FoM cohorts display a log normal distribution, where the mean value of $\ln(\Theta)$ for the PIB, LIB and SIB – All Materials, LIB and SIB – 2D Materials cohorts are 6.6E-12 $(m^2/s)$, 1.7E-11 $(m^2/s)$ and 5E-13 $(m^2/s)$ respectively.

Comparing these distributions, PIB rate performance falls between both LIB and SIB cohorts, LIB and SIB – All Materials show the highest mean values of FoM, whereas LIB and SIB – 2D materials shows far weaker rate performance than both groups. Although PIBs show a lesser mean value of FoM compared to the LIB and SIB – All materials cohort, these electrodes also show a larger spread of distribution. The standard deviation of the PIB distribution is greater than both the LIB and SIB cohorts. Due to this larger standard deviation, PIB electrodes show that while the mean FoM lies lesser than all LIB and SIB materials, some PIB electrodes can reach an equivalent rate performance to the highest rate performances displayed by LIB and SIB electrodes.

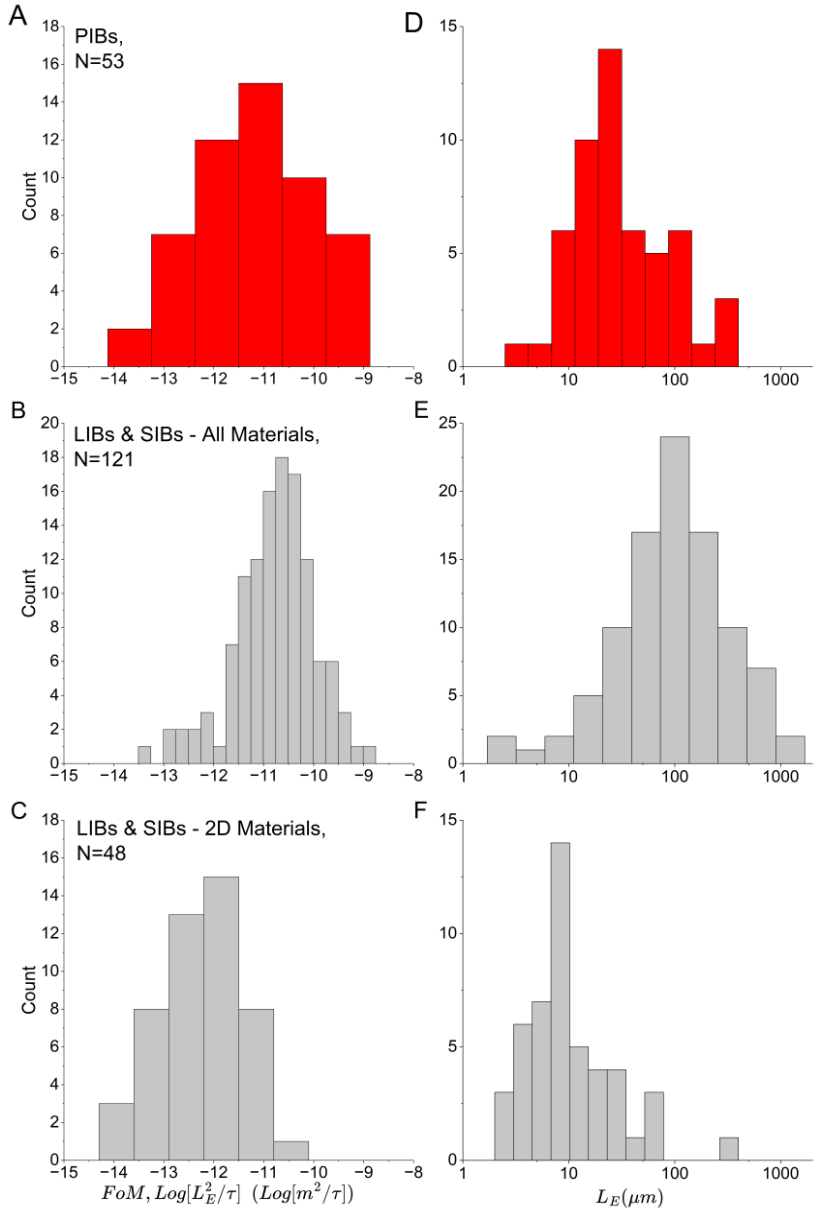

*Figure 3. Histograms comparing the figure of merit for rate performance (A-C) and the thickness of electrodes (D-F) between PIB electrodes (A,D), a large dataset consisting of LIB & SIB electrodes (B,E) and 2D material electrodes in lithium and sodium based chemistries (C,F). [17-49]*

Assessing this significant deviation within PIBs, there is potential the deviation is caused by the cohort representing very thin electrodes. As discussed earlier, the basis of our figure of merit is that at significant thicknesses $\tau$ approximately depends on $L_E^2$. However, we also note that prior to significant thicknesses $\tau$ depends approximately on the $L_E$ and/or constant terms, terms 4-7, as the $L_E^2$ terms become insignificant in comparison. In fact, at very small thicknesses $\tau$ will not be dependent on $L_E$ instead the constant parameters will be the only significant value and as thickness increases only after some less significant value of thickness will $\tau$ depend on $L_E$. This means that applying our figure of merit to very thin electrodes assigns them a disproportionately poor rate performance as it expects $\tau$ to decrease as with $L_E^2$ although this only occurs within proportion to $L_E$.

Due to this potential relationship between FoM and $L_E$ in PIBs, the comparison of FoM alone may not be fair, instead in cases where electrode thicknesses are particularly thin a plot of FoM versus electrode thickness can be used for an equitable comparison. To gauge whether the PIB sample's electrode thicknesses are particularly thin, Figure 3(D-F) shows the distribution of electrode thicknesses for (D) PIB materials, (E) LIB and SIB - all materials, and (F) LIB and SIB 2D materials, with $L_E$ ($\mu m$) on the x axes and the amount of electrodes, the count, on the y axes.

These plots show that the distribution of electrode thicknesses for the PIB cohort ranges over electrode thicknesses far thinner than that of the LIB and SIB - all materials. The PIB cohort show a mean $\ln(L_E)$ of 30 $\mu m$ as compared to a LIB and SIB - all materials with a mean of 91.2 $\mu m$. The thin nature of the PIB electrodes indicate investigation into a plot of FoM versus $L_E$ is worthy and comparison between distributions of FoM likely may not be fair to determine which group has better rate performance.

**Effect of Thickness**

Capacity and rate have been long known to be dependent on an electrode's thickness [50]. Having quantified low-rate specific capacity $Q_M$, characteristic time $\tau$, and high rate decay $n$, this provides an excellent opportunity to analyse the relationship of these properties to electrode thickness, $L_E$, in potassium-ion batteries. Below, Figure 4 shows plots of (A) $\tau$ (h) versus $L_E$ ($\mu$m), (B) n versus $L_E$ ($\mu$m), (C) $Q_m$ ($mAh/g$) versus $L_E$ ($\mu$m), and (D) $Q_M/Q/M_{Theoretical}$ versus $L_E$ ($\mu$m).

Figure 4A demonstrates that as electrodes become thinner, $\tau$ decreases and thus rate improves. As in our previous work assessing LIBs and SIBs, $\tau$ appears to scale with $L_E^2$ as expected. However, it is clear from this plot that there is more going on than simply following this trendline, there is a lot of scatter following different paths suggesting more at play. As a result, the nature of plot A is investigated and discussed further below. In Figure 4C specific capacity increases as electrodes become more thin. This relationship between capacity and thickness is highlighted in our plot by the datasets of the same material with varied thicknesses, marked with a hollow interior. While in Figure 4B, $n$ shows no dependence upon $L_E$ according to the plot, it shows scatter centred around n ~ 0.5. We note that these relationships mirror the relationships we have previously established for LIBs and SIBs and demonstrate evidence that our models and equations work appropriately with PIBs as with LIBs and SIBs. [17, 18]

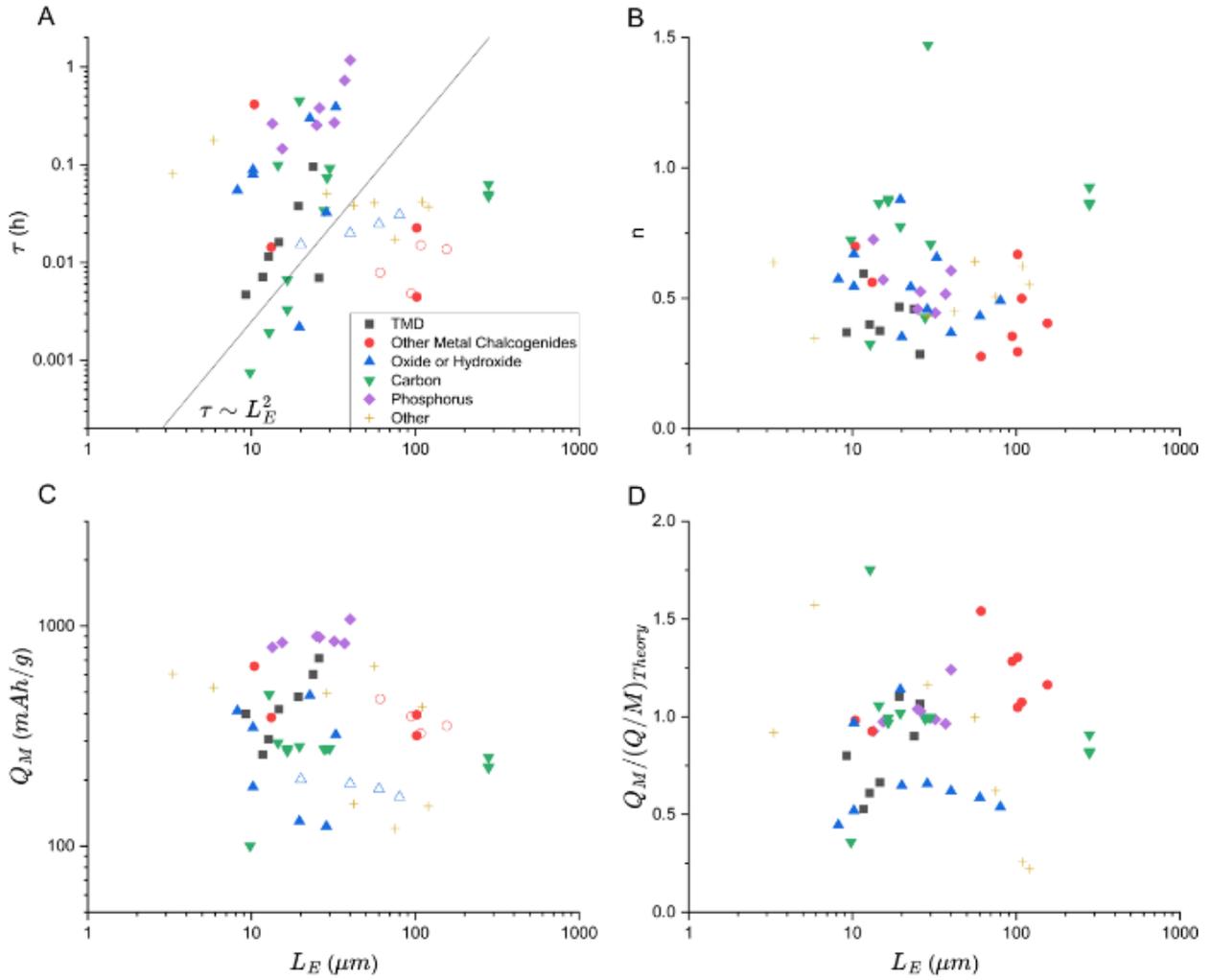

*Figure 4. Fitted parameters dependence on thickness. $L_E$ (μm) dependence of (A) time constant, $\tau$ ($h^{-1}$), (B) exponent, n, (C) low rate capacity, $Q_M$ (mAh/g), (D) $Q_M$ normalised to theoretical specific capacity. Open symbols relate to datasets of the same material where the thickness was varied, datasets like this were found within two papers, one series of datasets for the material category of oxides and hydroxides and another in other metal chalcogenides. The legend in plot (A) relates to all plots. [19-49]*

Additionally, evaluating Figure 4D, in different electrode chemistries, e.g. LIB, SIB, and PIB, certain active materials are known to hold working specific capacities poorer than their theoretical capacities, while others are known to outperform their theoretical capacities consistently. To determine patterns of specific capacity performance compared to theoretical capacity, $Q_M$ was normalised to its electrode's material's theoretical capacity, $Q/M_{Theoretical}$. Figure 4D shows a plot of $Q_M/Q/M_{Theoretical}$ versus $L_E$ (μm), demonstrating that most materials are performing within the region of their theoretical capacities with two notable exceptions. Other Metal Chalcogenides are routinely performing better by specific capacity than their theoretical capacity ought to allow, whereas Oxides and Hydroxides are consistently underperforming against their theoretical capacities.

Returning to the discussion of Figure 4A, to investigate the nature of what is driving the unique scatter, the effect of additional parameters on this plot was reviewed. It was found this plot's features could be broken down by the fraction of $\tau$ made up by the characteristic time associated with solid-state diffusion, $\tau_{SSD}$, which is derived from Equation 3 in a future section. The plots within Figure 5 break down these features within 3 regimes. Plot B represents $\tau_{SSD}/\tau \geq 0.99$, which highlights the datasets with values of $\tau$ above the trendline, the feature of the distribution which appears to follow a steeper relationship with $L_E^2$. Plot C represents $\tau_{SSD}/\tau < 0.99$, which highlights

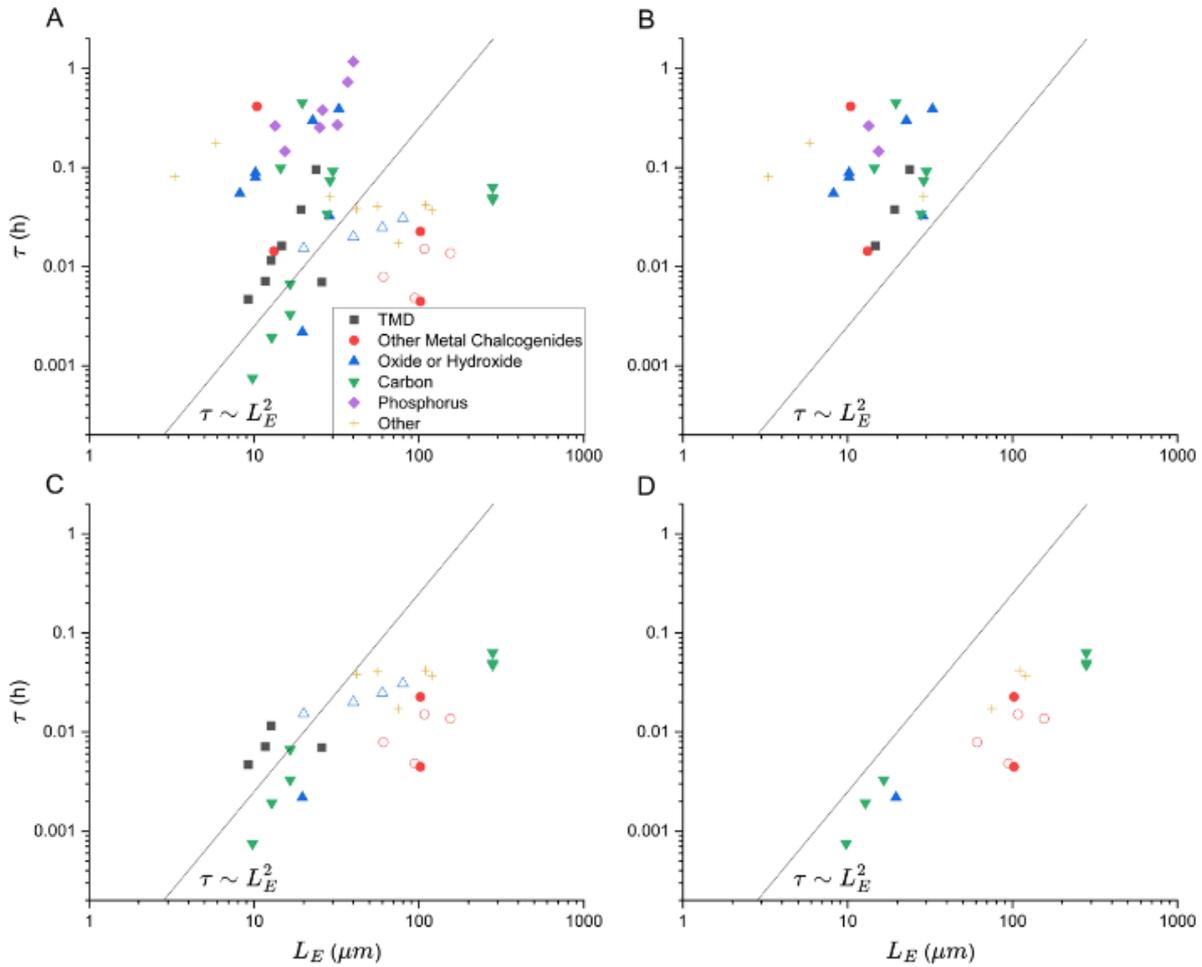

*Figure 5. Understanding the plot of $\tau(h^{-1})$ versus $L_E$ ($\mu m$). (A) Displays the dependence of $\tau$ on electrode thickness. The characteristic time, $\tau$ is the sum of characteristic times including the characteristic time associated with solid state diffusion, $\tau_{SSD}$, here portions of (A) the plot of $\tau$ versus $L_E$ are isolated based on the significance of $\tau_{SSD}$ within $\tau$. These plots isolate datasets where $\tau_{SSD}/\tau$ is greater than or equal to 0.99 (B), less than 0.99 (C), and less than or equal to 0.95 (D). Open symbols relate to datasets of the same material where the thickness was varied, datasets like this were found within two papers, one series of datasets for the material category of oxides and hydroxides and another in other metal chalcogenides. The legend in plot (A) relates to all plots. [19-49]*

the remaining data following more closely along, however not strictly following, the expected trend. While, plot D represents $\tau_{SSD}/\tau \leq 0.95$, datasets of more significant thicknesses, this group is offset below the trendline but strictly follow along its path, demonstrating the expected relationship scaling with $L_E^2$. From these relationships it is clear that the features and nature of the plot of $\tau$ versus $L_E$ is caused precisely due to a very high fraction of characteristic time which is made up by solid-state diffusion. The potential reasons that $\tau_{SSD}$ would represent such a high proportion of $\tau$ are that $\tau_{Diffusion}$ is particularly fast making $\tau_{SSD}$ most dominant, $\tau_{SSD}$ is extraordinarily large, or electrodes are unusually thin and the impact of the terms of Equation 2 dependant on $L_E$ and $L_E^2$ have yet to become dominant as proposed previously.

Figure 3 established that these electrodes appear to be particularly slim in comparison to the other electrodes, lending credibility to the notion that $L_E$ and $L_E^2$ terms are not yet dominant causing this multi featured nature of the plot of $\tau$ versus $L_E$. To determine more precisely if the source of the multi featured nature of Figure 4A is caused by very thin electrodes, Figure 6 shows the datasets where thickness has been varied alone.

For further analysis, a simplified quadratic of Equation 2, $\tau = aL_E^2 + bL_E + c$, was used to fit these datasets to determine a, b and c terms, which are visible within Figure 6. These datapoints appear to follow the early path of this quadratic where $L_E^2$ terms are less significant due to the extraordinarily thin nature of the electrodes studied. The path these data points trace shows that were the electrode data from more significant thicknesses these electrodes would strictly follow the trendline that is common among these $\tau$ versus $L_E$ plots. Thus it is clear from Figure 6, that the features and nature of Figure 4A are due to the electrodes being of a particularly slim nature. Additionally, as the electrodes are of a particularly slim nature, a fair

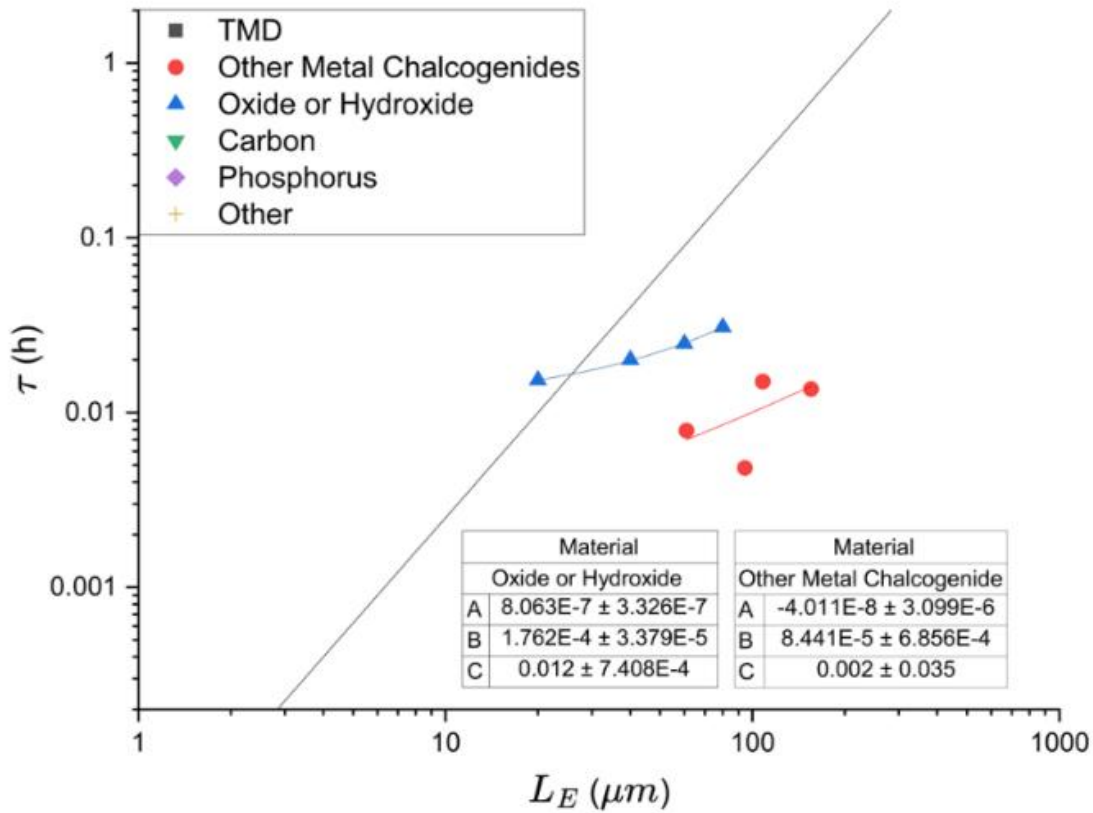

*Figure 6: Understanding the plot of $\tau$ ($h^{-1}$) versus $L_E$ (μm), demonstrating the effect of thickness on $\tau$. This plot displays the two sets of electrodes which had their thickness varied, where their other properties remained the same. Including line plots for both datasets based on simple quadratic of Equation 2. [25, 47]*

comparison cannot be made by directly comparing FoMs between this PIB cohort with others.

**Evaluating the Upper Limit of Rate Performance**

In other sections, the comparison of different electrode types' FoM distributions has proven unfair, as this sample of PIB electrodes are particularly slim and the FoM in very slim electrodes varies with $L_E$. However, there is a method which allows direct comparison of FoM distributions, even when one or each have particularly slim electrodes. By plotting inverse FoM versus $L_E$, the dependence of FoM's on $L_E$ can be seen. Although, while this trend cannot be fit to equations given the variance between electrode construction, bounds of the upper and lower limits can be fitted based on Equation 3. Assigning bounds allows each electrode type's upper and lower bounds to be compared equitably.

These bounds also allow us to understand rate performance of thicker electrode batteries. Which is consequential as the typical commercial battery electrode ranges in thickness between 50 $\mu$m – 300 $\mu$m, while this sample for PIBs has an average thickness of 30 $\mu$m. Typical electrodes also maintain an areal capacity greater than 2 $mAh/cm^2$, although no data in the PIB sample meets this criteria as seen within Figure 7C. Figure 7C shows a plot of areal capacity ($mAh/cm^2$) versus $L_E$ ($\mu$m) for the PIB cohort, this plot shows only two datasets of the sample exceed 1 $mAh/cm^2$. Considering this group of electrodes largely don't fit the electrode thickness of commercial electrodes and don't meet the areal capacity requirements, it is necessary to understand how thicker electrodes will perform in terms of rate for the practical use of these batteries.

The upper and lower bounds for $\tau/L_E^2$ versus $L_E$ were estimated based on both assumed reasonable values and by fitting some parameters. Assumed values are: $\sigma_{BL}$, bulk electrolyte ion conductivity of 1 S/M, $L_S$, separator thickness of 20 μm, $P_S$, separator porosity of 0.5, and $f$, upper and lower bound tortuosity factors of 0.5 and 1 respectively. Fitted parameters were: $Q_V$, upper and lower bound volumetric capacities were fitted based on their upper and lower values within the PIB cohort, $L_{AM}^2/D_{AM}$, upper and lower bound values of $L_{AM}^2/D_{AM}$ were fitted to the distribution based upon x axis translation, and $D_{BL}$, was fitted based upon the translation of both line plots along the y axis.

Figure 7B shows a plot of inverse FoM ($s/m^2$) versus $L_E$ ($\mu$m) for PIB electrodes. The plot demonstrates that all datasets reside within the range of thicknesses where FoM depends upon $L_E$ as expected. Included in this plot are fitted boundaries for this distribution. This distribution of inverse FoM vs $L_E$ displays a relationship inversely proportional to $L_E$, that is highlighted well by the datasets where $L_E$ has been varied. This relationship between inverse FoM and $L_E$ is consistent with Equation 2, which is evident from the plot as the slope of the fitted boundaries and the slope of the distribution match closely. Although a fitted boundary is displayed, it is clear from the distribution in Figure 7B a true lower limit has not yet been reached. The thicknesses of available electrodes have been too slim and none have been recorded at thicknesses where $L_E^2$ terms are overwhelmingly dominant. Therefore, this limit is an estimation based on current data and future electrodes may show that these batteries have a greater lower limit of rate performance than what is seen here.

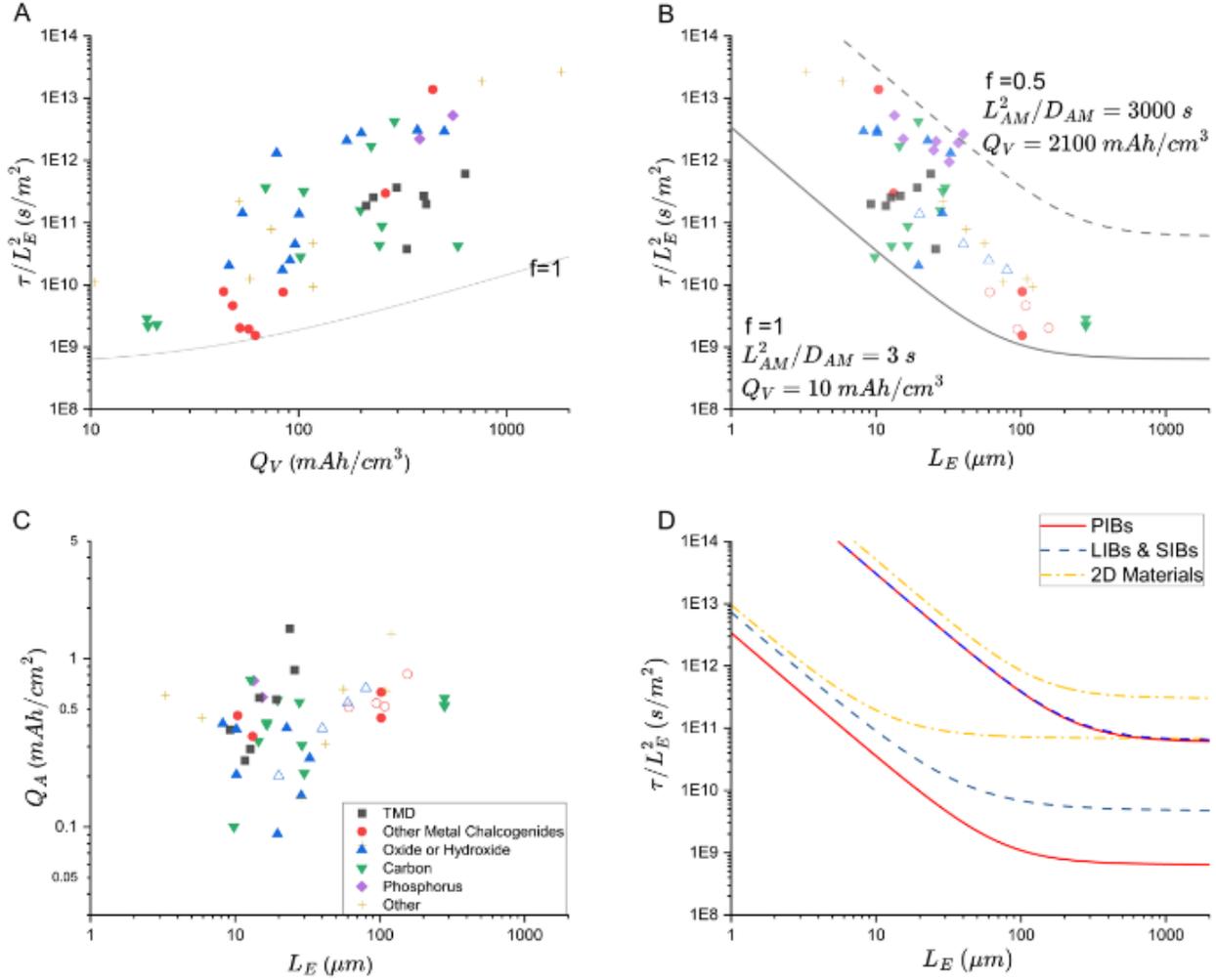

*Figure 7. (A) Inverse of the FoM, time constant normalised to the square of electrode thickness, $\tau/L_E^2$, plotted versus volumetric capacity of the electrode, $Q_V$, for PIB electrodes. The line in (A) is a plot of Equation 4, where f=1, $L_S$=20μm, $\sigma_{BL}=\sigma_{P,S}$=1 S/m and $D_{BL}$=$2 \times 10^{-9}$ $m^2/s$. The equation models the lower limit of $\tau/L_E^2$, an approximation which is valid for thick electrodes or short diffusion times. In which rate performance is only limited by ionic motion in the electrolyte within the porous interior of the electrode. The plotted line represents the case where the ionic conductivity and diffusivity within the pores are equal to their values in bulk electrolyte. (B) Plot of $\tau/L_E^2$ versus $L_E$, electrode thickness, for PIB electrodes. The lines represent upper(dashed) and lower(solid) limits of $\tau/L_E^2$ for a given $L_E$. These are plots of Equation 3 where $P_S$=0.5, $\sigma_{BL}$=1 S/m, and $D_{BL}$= $2 \times 10^{-9}$ $m^2/s$ and the parameters displayed within the plot. (C) Assesses the relationship between areal specific capacity (mAh/$cm^2$) and $L_E$ (μm), electrode thickness. (D) Compares the upper and lower limits of the PIB based electrodes with the LIB & SIB all material electrodes and 2D material electrodes established in a previous review. Open symbols relate to datasets of the same material where the thickness was varied, datasets like this were found within two papers, one series of datasets for the material category of oxides and hydroxides and another in other metal chalcogenides. The legend within the figure applies to plots (A,B). [17-49]*

Figure 7D demonstrates a fair comparison between all electrode types, a plot of inverse FoM ($s/m^2$) versus $L_E$ (μm), comparing the upper and lower limits of FoM for rate performance of PIB materials, LIB and SIB – All materials, LIB and SIB – 2D materials. This plot includes trendlines of lower limit and upper limit boundaries for all electrode types, where those for the LIB and SIB datasets were garnered from our previous work [17, 18]. This plot demonstrates that PIB electrodes show a superior lower and upper limit of rate performance to either LIB and SIB dataset.

The source of the increased performance becomes clear upon comparing the parameters used between electrode types within Equation 3. For the lower bounds ascribing the best rate performance, parameters for PIBs, LIB and SIB - All Materials, and LIB and SIB – 2D Materials respectively are: $f$=1, 1, 0.1, $L_S$=20 $\mu m$, 25 $\mu m$, 25 $\mu m$, $\sigma_{BL}$=1 $S/m$, 1 $S/m$, 1 $S/m$, $Q_V$=10 $mAh/cm^3$, 100 $mAh/cm^3$, 250 $mAh/cm^3$, $L_{AM}^2/D_{AM}$=3 s, 3 s, 5 s, and $D_{BL}$= $2 \times 10^{-9}$ $m^2/s$, $3 \times 10^{-10}$ $m^2/s$, $3 \times 10^{-10}$ $m^2/s$. Diffusion within bulk electrolyte, $D_{BL}$, is far superior within PIBs, improving on LIB and SIB electrodes by almost an order of magnitude. These parameters demonstrate that the diffusion of potassium ions is superior to lithium and sodium ions in electrolyte.

This estimate for the diffusion coefficient for potassium ions in bulk electrolyte is in fact consistent with observed diffusion coefficients in research. Research papers show the diffusion coefficient for potassium ions in liquid electrolytes ranges from $1 \times 10^{-9}$ to $3 \times 10^{-9}$ $m^2/s$, while LIBs and SIBs share an almost identical range for their respective diffusion coefficients of $0.5 \times 10^{-10}$ to $4 \times 10^{-10}$ $m^2/s$. In fact, research papers comparing the diffusion of potassium with lithium show this same relationship with potassium outperforming in terms of diffusion. Surprisingly, despite the increased size and mass of potassium ions compared to lithium and sodium ions, potassium shows the best diffusion of the three ions in liquid electrolyte. [51-56]

Additionally, to address the final plot in Figure 7, considering previous work [18] where 2D materials were found to be limited due to their tortuosity factor, it was deemed valuable to determine if there were any limitations due to the tortuosity and porosity of PIB electrodes. This previous work established a simplified equation, derived from Equation 3, which allows a boundary to be created representing a minimum $\tau/L_E^2$ based upon $f$, the tortuosity factor within the electrode. This allows a minimum tortuosity, or maximum tortuosity factor, to be fitted to the dataset of a battery cohort to determine whether there are tortuosity/porosity limitations. The equation derives from the first two terms of Equation 3, the terms representing electrical conductivity and ion diffusion within the porous interior of the electrode, these determine a lower limit for the FoM:

$$\left(\frac{\tau}{L_E^2}\right)_{min} \approx \frac{14 Q_V}{\sigma_{BL} f} + \frac{1}{D_{BL} f} \qquad (4)$$

based upon previous assumed parameter values and by fitting $f$ to the cohort. Figure 7A demonstrates a plot of inverse FoM ($s/m^2$) versus $Q_V$ ($mAh/cm^3$), along with a fitted trendline of minimum tortuosity. In plot A, the minimum tortuosity factor is unity, the best achievable value, similar to those of LIB and SIB all materials suggesting no limitation from porosity or tortuosity.

**Evaluating Components of Solid-State Diffusion**

It is now established that PIBs show greater ion diffusivity compared to LIB and SIB electrodes, however rate performance and $\tau$ are not solely dependent on diffusivity within bulk electrolyte. Rate is also influenced by electrical limitations and the limitations of solid-state diffusivity within active materials. While electrical limitations have already been investigated and discussed, solid-state diffusivity limitations have not.

Characteristic time associated with solid-state diffusivity is an important parameter as it determines the most rapid rate performance of the electrode. The fastest rate performance, lowest values of $\tau$, occur where electrodes are particularly slim and diffusivity within the bulk electrolyte does not impact rate. By understanding characteristic time associated with solid-state diffusivity, we can understand the absolute limits of rate within PIB electrodes.

Previously, we determined that by rearranging Equation 3, we can determine the characteristic time associated with solid-state diffusion, $\tau_{SSD}$. As solid-state diffusivity is associated with active particle size and solid-state diffusivity, we associated term 6 of Equation 2, $L_{AM}^2/D_{AM}$, with $\tau_{SSD}$. Solving Equation 3 for this term we yield the equation:

$$\tau_{SSD} = \tau - L_E^2 \left[ \frac{14Q_V}{\sigma_{BL}f} + \frac{1}{D_{BL}f} + \frac{28Q_V L_S/L_E}{P_S \sigma_{BL}} + \frac{L_S^2/L_E^2}{P_S D_{BL}} \right] \quad (4)$$

And, as we determined the upper and lower bounds for $f$ previously, we used the average value to evaluate each electrode's $\tau_{SSD}$.

Using Equation 4, the proportion of $\tau$ made up by the characteristic time associated with solid-state diffusion, $\tau_{SSD}$, was calculated for all electrodes. The distribution of $\tau_{SSD}$ for PIB, LIB and SIB – all material, and LIB and SIB – 2D material electrodes can be seen within Figure 8A,C,E below. The plots of Figure 8A,C,E show lognormal distributions similar to the distribution of $\tau$ in Figure 3. The mean values of PIB, LIB and SIB – All Material, and LIB and SIB – 2D material electrodes are 0.0272 h, 0.061 h, and 0.055 h. Comparing the distributions of Figure 8A,C,E, within solid-state diffusion PIBs maintain their superiority in rate performance, even outperforming 2D materials that would be expected to have superior solid-state diffusion due to restriction of one dimension to virtually dimensionless size. Although, while PIBs outperform both LIB and SIB datasets in terms of mean $\tau_{SSD}$, both LIB and SIB datasets maintain some materials with values of $\tau_{SSD}$ at or faster than the PIB cohort's fastest rate.

While it is important to understand how the distribution of this parameter of rate compares to other electrode types, it is also important to understand how it affects $\tau$ as thickness increases. The expected composition of $\tau$ as electrode thickness increases is: in slim electrodes the composition of $\tau$ is wholly made up of $\tau_{SSD}$ while as thickness increases limitations of diffusion within bulk electrolyte and electrode pores begin to be more prevalent until at particularly high thicknesses $\tau_{SSD}$ becomes an insignificant limitation. Were this behaviour witnessed, it would provide more evidence that the equations for rate performance effectively model electrode behaviour in PIBs. While it additionally would provide insight as to the veracity of the superior rate performance of PIBs, because as diffusion within bulk electrolyte becomes improved it would be expected that its characteristic time is lesser and thus that the $\tau_{SSD}$ proportion of $\tau$ is larger for greater thicknesses.

The proportion of $\tau$ made up of solid-state diffusion can be gleamed by normalising $\tau_{SSD}$ to $\tau$. Using this parameter, $\tau_{SSD}/\tau$ is plotted against electrode thickness ($\mu m$) for PIBs, LIB and SIB – All Materials, and LIB and SIB – 2D materials in Figure 8B,D,F. These plots show the expected behaviour, within slim electrodes the proportion of characteristic time is almost completely made up of solid-state diffusion time, whereas as electrodes become more thick this proportion of $\tau$ made up of $\tau_{SSD}$ decays in a logarithmic fashion, with scatter caused by the variance between materials and electrode construction. Comparing the plots B,D,F of Figure 8 it is clear the decay of the proportion of $\tau_{ssd}$ to $\tau$ begins much later in PIBs to LIB and SIB – All Materials and later again from LIB and SIB – All Materials to LIB and SIB – 2D Materials. This analysis of the characteristic time associated with solid-state diffusion reaffirms the success of our semi-empirical model in addition to the finding that the diffusivity of ions in bulk electrolyte is superior in PIBs.

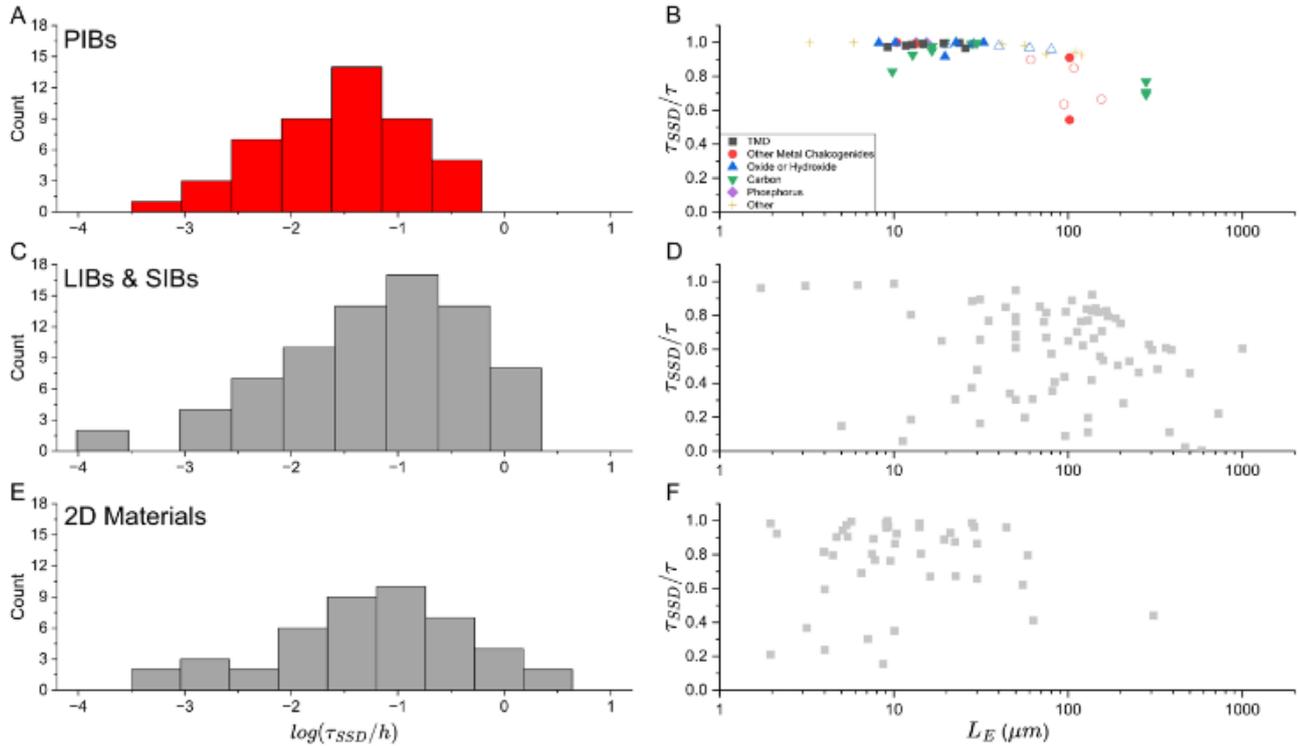

*Figure 8. (A, C, E) Histograms comparing estimated solid-state diffusion time, $\tau_{SSD}$ ($\tau_{SSD}=L_E^2/D_{AM}$) for (A) PIB based electrodes, (B) LIB & SIB based electrodes and (C) 2D material based LIB& SIB based electrodes. (B, D, F) Plots of $\tau_{SSD}/\tau$ versus $L_E$ (µm) comparing active materials for (B) PIBs, (D) LIBs & SIBs and (F) 2D based materials for LIBs & SIBs. $\tau_{SSD}$ calculated based on Equation 4, using estimated values $\sigma_{BL}=1$ S/m, $D_{BL}=2\times10^{-9}$ $m^2/s$, $L_S=20$ µm, $P_S=0.5$, $f=0.75$ and deriving values for $Q_V$ based on $Q_M$, mass loading and thickness. Open symbols relate to datasets of the same material where the thickness was varied, datasets like this were found within two papers, one series of datasets for the material category of oxides and hydroxides and another in other metal chalcogenides. [17-49]*

**Conclusions**

The aims of this work were to evaluate the applicability of our semi-empirical model on PIBs, to evaluate the performance of PIBs and find trends that might help inform future improvement in performance, and to compare the performance of PIBs against LIBs and SIBs using the characteristic parameters determined by our model. Across these objectives, this work has demonstrated the suitability of our methods for use with emerging battery chemistries, highlighted novel insights into PIB performance and opportunities in their performance characteristics, and positioned PIB performance in terms of the current and emerging alkali-ion energy storage systems.

The first objective of this research was to determine the success of our semi-empirical model on PIBs, given it was first proposed as a solution for all batteries but only reviewed using the characterisation of LIBs and SIBs. The findings demonstrate that our model is indeed capable of fitting PIBs with a great degree of accuracy. While some inaccuracy was found in the fitting of parameters of n, it was found this could be mitigated by increasing the availability of high-rate data, and the robustness of the fitting demonstrates that PIBs are consistent with the relationships of our model and equations. This finding is significant because it demonstrates that the experimental performance of PIBs can be quantified to perform more rigorous analysis and comparison, enabling experimental researchers in this field to use this simple model to identify novel performance more readily.

The second objective of this research was to assess the performance of PIBs and find trends that might help inform future improvement in performance, because by quantifying the PIB experimental data

PIB performance overall can be assessed through this representative sample. Within this investigation numerous trends were identified. One overall trend found that many PIB electrodes in research suffered from electrical limitations, something which is known and was found in previous work to be mitigated by the use of greater amounts of conductive additive within the construction of the electrode. Other trends regarded the comparison of the performance of the different PIB electrode active materials. A key relationship previously established was that capacity is inversely proportional to rate performance. Although, the phosphorus, transition metal dichalcogenide, and other metal chalcogenide active materials demonstrated better rate performance and specific capacity relative to this correlation in PIBs. This highlights that phosphorus, transition metal dichalcogenides, and other metal dichalcogenides may demonstrate uniquely beneficial properties which would benefit PIBs and warrant greater focus within research.

The final objective of this research was to use the large cohort of fit characteristic performance parameters and compare the performance of PIBs with the previously assessed LIB and SIB cohorts, as this quantification of PIBs establishes the greatest opportunity of performance comparison among these batteries. What was found revealed a nuanced positioning of PIB performance. While LIBs and SIBs dominated in terms of capacity performance, PIBs demonstrate greater performance in terms of charge/discharge rate capability. Although the true upper limit of the rate performance of PIBs could not be identified, due to the lack of PIB electrodes with significant thicknesses (>100 μm), the analysis completed demonstrates a greater upper limit of rate performance for PIBs compared to both LIB and SIB cohorts. This demonstrates a potential unique performance characteristic for PIBs among these alkali-ion batteries and potential for the applicability of these batteries commercially.

**Methods**

Capacity versus rate data was gathered from published research papers via the "digitizer" function in Origin. The Charge/discharge rate is typically expressed using current or C-rate, these are converted to rate, R, using the equations given in ref [17]. Fitting was executed in Origin Pro software (Origin 2024), using a custom fitting function based on Equation 3, within the "non-linear curve fit" function. All fits and additional data is provided in the SI.

However, in many instances necessary data was not provided plainly within the text of published works Although, both active material loading (mg/cm$^2$) and the proportions of active material versus conductive additive and binder typically are, the electrode thickness is rarely provided. This is unfortunate, because as Equation 2 shows thickness has a critical effect on rate performance. To facilitate rate analysis, in many cases we were forced to estimate electrode thickness (See SI). This was performed using: (i) the total mass loading and mass fraction of active material; (ii) the densities of active material and binder/additive combination; and (iii) the electrode porosity. Where the (i) parameters are not available analysis cannot be completed. The (ii) parameters can often be estimated with reasonable accuracy. However, the porosity (iii) is seldom provided and in these cases we were forced to estimate porosity. Estimated electrode porosity was given as $P = 0.5$. This is justifiable for electrodes as the ideal porosity lies between 0.4 and 0.6[57-59], assuming actual porosity lies within this range produces an error of 20%. To yield an error in electrode thickness, assuming the mass loading error is ~10%, yields a 30% error which is acceptable given the very broad range of distribution for values of $L_E^2/\tau$. We note that the set of papers examined, cannot be considered complete, as so many papers which report rate performance have not provided sufficient information to perform our analysis.

The datasets in Figure 3, Figure 7, and Figure 8 demonstrating lithium-ion batteries and sodium-ion batteries, labelled the LIB and SIB – All materials and LIB and SIB – 2D Materials, are extracted from refs [17, 18].